\documentclass[pdflatex,sn-mathphys-num]{sn-jnl}%

\usepackage{graphicx}%
\usepackage{multirow}%
\usepackage{amssymb,amsfonts}%
\usepackage{amsthm}%
\theoremstyle{definition} 
\usepackage{float}
\usepackage{mathrsfs}%
\usepackage{xcolor}%
\usepackage{textcomp}%
\usepackage{longtable}
\usepackage{manyfoot}%
\usepackage{setspace}
\usepackage{booktabs}%
\usepackage{tabularx} 
\usepackage{algorithm}%
\usepackage{algorithmicx}%
\usepackage{algpseudocode}%
\usepackage{listings}%
\usepackage{enumitem} 
\usepackage{bm}
\usepackage{multicol}
\usepackage{colortbl}
\usepackage{adjustbox}
\usepackage{subcaption}
\usepackage{algcompatible}
\usepackage{amssymb}
\usepackage{ulem}
\usepackage{xr}
\usepackage{hyperref}
\usepackage{ragged2e}
\usepackage[tbtags]{amsmath}
\usepackage{anyfontsize}
\usepackage{makecell}

\theoremstyle{thmstyleone}

\theoremstyle{thmstyletwo}%
\newtheorem{remark}{Remark}%

\theoremstyle{thmstylethree}%
\newtheorem{definition}{Definition}%

\begin{document}

\title[Article Title]{Nonparametric framework for the definition, adaptive detection and probabilistic interpretation of outliers}

\author*[1]{\fnm{Tiziano} \sur{Iannaccio}}\email{tiziano.iannaccio@uniroma1.it}

\author[1]{\fnm{Maurizio} \sur{Vichi}}\email{maurizio.vichi@uniroma1.it}

\affil[1]{\orgdiv{Department of Statistical Sciences}, \orgname{Sapienza University of Rome}, \orgaddress{\street{Piazzale Aldo Moro 5}, \city{Rome}, \postcode{00185}, \country{Italy}}}

\abstract{Outlier detection is a fundamental challenge in data processing, with critical implications for robustness across statistical modeling, machine learning and exploratory data analysis. However, existing proposals rarely offer a universal, domain-agnostic definition of an outlier, often relying on heuristic trimming quotas that lack a statistical interpretation. To address this, we propose a nonparametric framework built on a pseudo-isolation outlier score. This score enables a formal, probabilistic definition of an anomaly tied to a false-alarm rate $\alpha$, which extends into a rigorous geometric classification of internal and external outliers. We show that this mechanism seamlessly embeds into any objective-based clustering framework to identify cluster-specific outliers. Here, we integrate it into $K$-means to create ODK-means. The inferential capabilities and topological properties of this framework are explored both theoretically through formal propositions and empirically through extensive simulations and methodological tutorials, highlighting the practical actionability and intuitive appeal of the proposed outlier detection logic.}

\keywords{robust clustering, Outlier detection, Nonparametric inference, Anomaly classification}

\maketitle

\section{Introduction}\label{sec1:intro}

Outliers are generally considered as univariate or multivariate units that occupy atypical or sparsely populated regions of the data space. As noted by Knorr \& Ng (1997) \cite{knorr1997unified}, outliers can arise due to various reasons, such as measurement errors or the joint existence of different underlying processes. However, Hodge \& Austin (2004) \cite{hodge2004survey} specify that outliers may also carry valuable insights about rare events or anomalies that warrant investigation. The search for a unique, rigorous definition of anomaly is a central, ongoing challenge in outlier analysis. This issue is reflected in the spectrum of closely related characterizations offered across the literature over the past several decades. Unfortunately, most characterizations represent (helpful) descriptive guidelines that cannot be directly translated into a detection algorithm: what qualifies as an outlier shifts between datasets, domains and analytical goals.

The outlier status is typically viewed as binary. However, Breunig et al. (2000) \cite{breunig2000lof} argue that it is more informative to assign each observation a degree of "outlier-ness", such as the Local Outlier Factor (LOF), which is evaluated as the ratio of each item's isolation to its neighborhood's average isolation. Building upon this perspective, we introduce a standalone, distribution-agnostic outlier detection framework. Our proposal turns the empirical distribution of an isolation-based outlier score into a formal, probabilistic definition of an anomaly linked to a false-alarm rate $\alpha$. This detection mechanism formalizes the already established categorization of anomalies into internal and external outliers. Furthermore, by evaluating the minimum false-alarm rate $\alpha^\star$ required to flag a given observation, our framework unlocks a rich inferential background and reveals three topological properties: (i) a baseline region with no anomalies, (ii) intervals of stable outlier counts and (iii) the natural emergence of outliers in discrete groups.

In heterogeneous environments, anomaly detection and cluster estimation are mutually dependent problems. For this reason, Rousseeuw \& Van Zomeren (1990) \cite{rousseeuw1990unmasking} discuss the need to robustify clustering algorithms against outliers. Consequently, robust clustering literature frequently integrates outlier detection directly into the partitioning process. Among these proposals, most are extensions of the well-known $K$-means methodology (MacQueen 1967 \cite{macqueen1967multivariate}, Lloyd 1982 \cite{lloyd1982least}). Trimmed $K$-Means (1997) \cite{cuesta1997trimmed} represents a key development in this domain, enhancing resilience to outliers by iteratively trimming a pre-specified proportion $\hat{p}$ of the data points based on their distance to cluster centroids. This seminal work established an entire branch of literature dedicated to analyzing and refining its robust properties, extending from Garcia-Escudero \& Gordaliza (1999) \cite{garcia1999robustness} to recent advances such as Dorabiala et al. (2022) \cite{dorabiala2022robust}. \textit{(For a more comprehensive review of traditional and contemporary outlier literature, see Section S1 of the Supplementary Material.)}

While most detection techniques require fixing $\hat{p}$ prior to estimation, our probabilistic module can be seamlessly embedded into any objective-based clustering framework to achieve adaptive trimming. In doing so, the induced geometric taxonomy is extended to include cluster-specific outliers. In this work, we integrate our detection engine specifically into $K$-means, resulting in the Outlier Detection $K$-means (ODK-means) algorithm. Choosing the well-established $K$-means ensures that the primary focus remains on the inferential properties of the outlier detection framework, while allowing for a direct comparison against established robust benchmarks.

The remainder of this paper is organized as follows. Section~\ref{sec2:methodologies} formalizes the proposed outlier detection framework, while Section~\ref{sec3:background} provides its methodological background. Section~\ref{sec4:inference} analyzes the inferential and topological properties induced by the framework. Section~\ref{sec5:algorithms} details the algorithmic implementation and computational complexity. The empirical performance is evaluated through the simulation study in Section~\ref{sec6:simulations} and illustrated via a methodological tutorial in Section~\ref{sec7:tutorial}. Finally, Section~\ref{sec8:discussion} offers concluding remarks.

\section{Methodologies}\label{sec2:methodologies}

\subsection{Outlier threshold}
\label{sec21:outlier_threshold}
Let $N_i(h,l)$ be the set of units ranked $h$-th to $l$-th in terms of their proximity to unit $i$. For each $i$, we define its (Mahalanobis) \textit{pseudo-isolation} as

\begin{equation}
    \label{eq1:pseudo_isolation}
    y_i(h,l)=\sum_{j \in N_i(h,l)}||\mathbf{x}_i-\mathbf{x}_j||^2_{\hat\Sigma^{-1}}.
\end{equation}

\noindent The neighborhood ranks ($h,l$) will henceforth be omitted from the notation of $y_i$. Let $\tilde{\mathbf{Y}}=(y_{(1)},\dots,y_{(N)})'$ be the vector of pseudo-isolations corresponding to the sample at hand, arranged in non-decreasing order. For each $m \in \{1,\dots,N\}$ evaluate:

\begin{equation}
    \label{eq2:running_stats}
    \mu_m = \frac{1}{m}\sum_{i=1}^m y_{(i)}, 
\qquad
\sigma_m = \sqrt{\frac{1}{m-1}\sum_{i=1}^m \bigl(y_{(i)}-\mu_m\bigr)^2}
\end{equation}

\noindent and the candidate \textit{Cantelli-based} outlier threshold for a fixed $0 \leq\alpha\leq1$:
\begin{equation}
    T_{\alpha,m} = \mu_m + \sigma_m\sqrt{\frac{1}{\alpha}-1}.
    \label{eq3:runningthreshold}
\end{equation}

\noindent Let the optimal order $m^\star\in\{1,\dots,N\}$ be the solution of the following problem:
\begin{equation}
    \label{eq4:minT}
    \min_{m \in \{1,\dots,N\}} \quad  T_{\alpha,m} 
\end{equation}
such that:
\begin{equation}
    \label{eq5:minT_constraint}
    y_{(m)} \le T_{\alpha,m} < y_{(m+1)},
\end{equation}
\noindent where $y_{(N+1)} := +\infty$. Finally, let $T_\alpha^\star := T_{\alpha,m^\star}$ be the outlier threshold at level $\alpha$.
\subsection{Outlier Definition and Categorization}
\label{sec22:definitions}

The above setup allows us to give a general, domain agnostic and probabilistic definition and of outlier, together with a categorization in three classes of anomaly. 

\begin{definition} [Outlier at level $\alpha$]
    \label{def1:outlier_regular}
    \noindent Let $T_\alpha^\star$ be the outlier threshold at level $\alpha$ and $y_i$ be the pseudo-isolation of the multivariate unit $i$. If the following condition holds:
\begin{equation}
    \label{eq6:outlier_condition}
    y_i\geq T_{\alpha}^\star;
\end{equation}
unit $i$ is an outlier at level $\alpha$. Otherwise, unit $i$ is a regular observation at level $\alpha$.
\end{definition}

\noindent Let $\mathbf{S}$ be a set of $N$ data points. The convex hull of $\mathbf{S}$ is defined as:
\begin{equation}
\label{eq7:convhull}
    \operatorname{conv}(\mathbf{S}) 
    \;=\; \left\{ 
        \sum_{i=1}^N \lambda_i \mathbf{x}_i 
        \;\mid\; \mathbf{x}_i \in \mathbf{S}, \lambda_i \geq 0,\ \sum_{i=1}^N \lambda_i = 1
    \right\}.
\end{equation}

\noindent We then define $\mathbf{S}^\star=\{\mathbf{x}_i \in \mathbf{S} : y_i< T_{\alpha}^\star\}$ as the set of regular observations (null sample) and conv($\mathbf{S^\star}$) as the convex hull of the null sample. 

\begin{definition} [External Outlier at level $\alpha$]
    \noindent A multivariate unit $i$ is an External Outlier at level $\alpha$ if the following conditions hold:
    \begin{itemize}
        \item $i$ is an outlier at level $\alpha$
        \item $\mathbf{x}_i \notin \text{conv}(\mathbf{S^\star})$
    \end{itemize}
    \label{def2:external}
\end{definition}

\begin{definition} [Internal Outlier at level $\alpha$]
    \noindent A multivariate unit $i$ is an Internal Outlier at level $\alpha$ if the following conditions hold:
    \begin{itemize}
        \item $i$ is an outlier at level $\alpha$
        \item $\mathbf{x}_i \in \text{conv}(\mathbf{S^\star})$
    \end{itemize}
    \label{def3:internal}
\end{definition}

\noindent
The evaluation of $T_{\alpha}^\star$ and the detection of both external and internal outliers can be used \textit{independently} from the clustering process. However, the presence of internal outliers may suggest heterogeneity within the data. Therefore, integrating the procedure with a clustering algorithm enables the identification of \textit{cluster-specific outliers}. In order to do so, the procedure evaluates $K$ separate thresholds ${}_k T_{\alpha}^\star$: one for each cluster. Then, outlier detection is performed separately for each cluster. This yields the set $\mathbf{S}^\star_k$ of regular data in cluster $k$ (null sample in cluster $k$) and its convex hull conv($\mathbf{S}^\star_k$). 

\begin{definition} [Cluster-specific Outlier at level $\alpha$]
    \noindent A multivariate unit $i$ is a Cluster-specific Outlier at level $\alpha$ if the following conditions hold:
    \begin{itemize}
        \item $i$ is an outlier at level $\alpha$
        \item $\mathbf{x}_i \in \text{conv}(\mathbf{S}^\star_k)$
    \end{itemize}
    \label{def4:cluster_specific}
\end{definition}

\begin{remark}
    \label{rem1:cluster_internal_relation}
    A Cluster-specific outlier is always an Internal outlier, but not vice versa. This is because $\text{conv}(\mathbf{S}^\star_k)\subseteq \text{conv}(\mathbf{S}^\star)$ for all $k=1,...,K$. 
\end{remark}

\begin{figure}[H]
    \centering
    \includegraphics[width=0.85\textwidth]{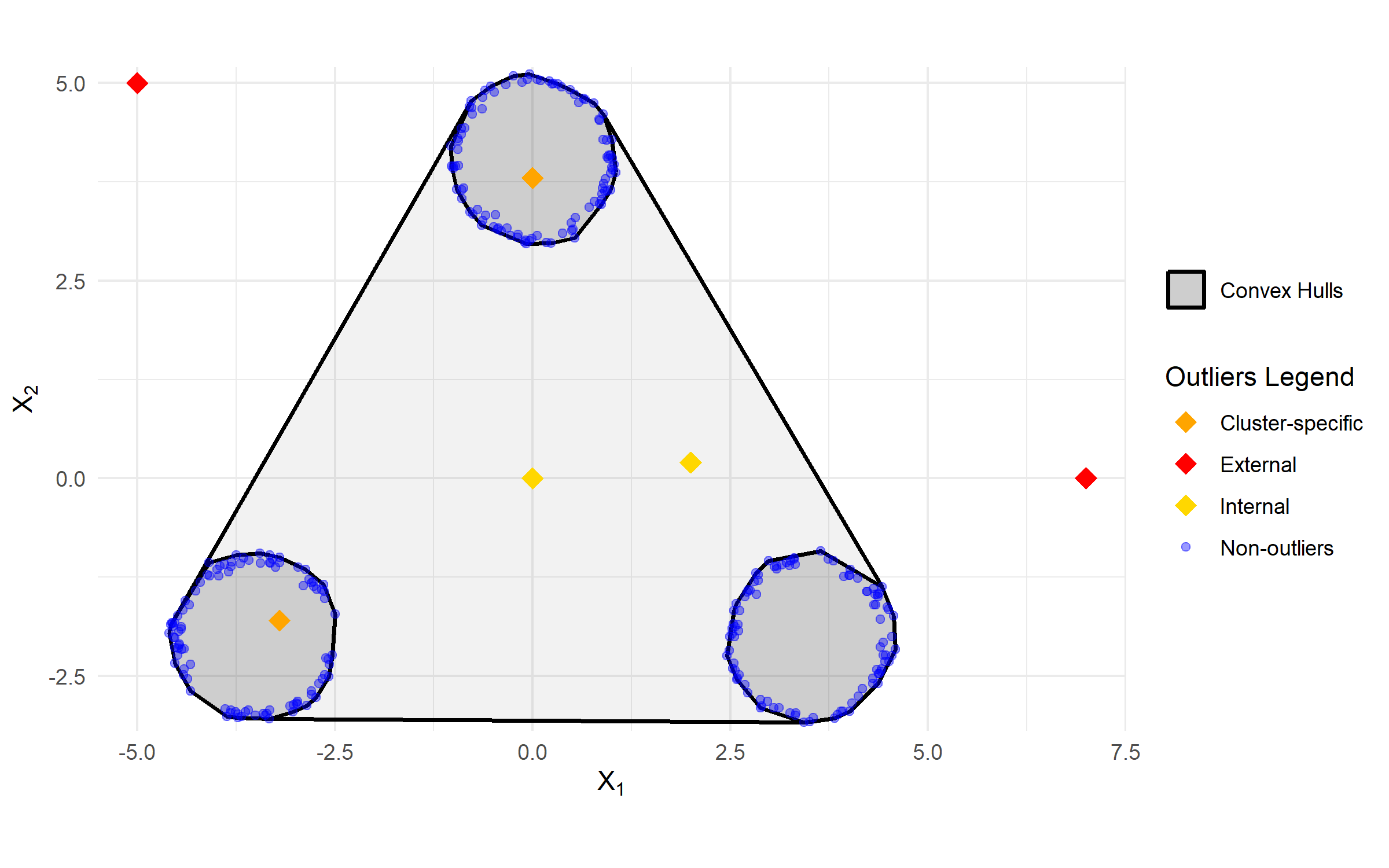}
    \caption{Illustration of the geometric taxonomy of multivariate outliers. An external outlier (red diamond) falls outside $\operatorname{conv}(\mathbf{S}^\star)$; a non-cluster-specific internal outlier (yellow diamond) resides inside $\operatorname{conv}(\mathbf{S}^\star)$ but outside any local hull; a cluster-specific outlier (orange diamond) falls within its corresponding local hull $\operatorname{conv}(\mathbf{S}^\star_k)$.}
    \label{fig1:taxonomy}
\end{figure}

\noindent To complement the rigorous formalizations above, Figure~\ref{fig1:taxonomy} offers an intuitive visual distinction between the three outlier categorizations in a simplified 2D scenario.

\noindent \textit{(We refer the reader to Section S2 of the supplementary material, which maps our methodology onto the well-established framework of cell-wise and case-wise outliers.)}

\subsection{ODK-means}
\label{sec23:odkm}
An intuitive procedure that simultaneously performs clustering and outlier detection can be modeled as follows: 
\begin{equation}
    \begin{aligned}
    \min_{u_{ik},\bar{\mathbf{x}}_k} & \sum_{i=1}^N\sum_{k=1}^K w_{i} u_{ik} \|\mathbf{x}_i - \bar{\mathbf{x}}_k\|^2_2\\
        s.t.&\\
    & u_{ik} \in \{0,1\}\ \ \forall i=1,\ldots,N;\ \forall k=1,\ldots,K;  \\ 
    & \sum_{k=1}^K u_{ik} = 1\ \ \forall i=1,\ldots,N.
    \end{aligned}
    \label{eq8:odk}
\end{equation}

\begin{remark}
   \label{rem2:weighted_kmeans}
   When the proposed outlier detection module is integrated into an objective-based clustering framework, the resulting model naturally translates to a weighted extension of the underlying objective. In particular, Problem~\ref{eq8:odk} is an instance of Weighted $K$-means.
\end{remark}

\noindent The contribution of each unit $i$ within its specific cluster $k$ can be weighted as follows:

\begin{equation}
    \label{eq9:weight_scheme}
    w_i = 
    \begin{cases} 
    1 & \text{if } y_i < {}_k T_{\alpha}^\star \\
    q\frac{{}_k T_{\alpha}^\star}{y_i} & \text{if } y_i \geq {}_k T_{\alpha}^\star\quad 
    \end{cases}
\end{equation}
\noindent The $hard\ trimming$ approach ($q=0$) is suggested to prevent extreme values from biasing the cluster centroids. The $soft\ trimming$ approach ($0<q\leq\frac{y_i}{{}_k T_{\alpha}^\star}$) is preferred in case of domain expertise: by tuning $q$, users can leverage their a priori knowledge of the system to calibrate the influence of anomalies. Weights are then normalized.

\section{Methodological background}
\label{sec3:background}
\subsection{Pseudo-Isolation as a Robust Outlier Score}\label{sec31:outlier_score}
Given the arbitrary shapes of multivariate data, non-parametric procedures provide a robust baseline for outlier detection by avoiding distributional assumptions. This idea dates at least back to Amidan et al. (2005) \cite{amidan2005data}, who introduced a two-stage outlier-trimming routine based on Chebyshev’s inequality. Pang et al. (2018) \cite{pang2018learning} refined this approach by defining a positive outlier score $Y$, where higher values of $Y$ correspond to a more anomalous behavior. This detail allows the outlier threshold to be evaluated via the Cantelli (or unilateral Chebyshev) inequality, recalled here: let $Y$ be a random variable (e.g., the one generating pseudo-isolations) with finites mean $\mu$ and variance $\sigma^2>0$. Then, for any $c>0$:
\begin{equation}
    P(Y - \mu \geq c\sigma) \leq \frac{1}{1 + c^2}.
    \label{eq10:cantelli_base}
\end{equation}

\noindent These proposals motivate both our metric choice and threshold design. The pseudo-isolation in Eq.~\eqref{eq1:pseudo_isolation} is indeed a positive outlier score that adapts the nearest-neighbor metric of Pang et al. to our framework through three key modifications. First, rather than estimating local isolation over random subsamples, we preserve the original spatial ordering of the complete sample. Second, rather than measuring isolation relative to the closest neighbors of the reference sample, we start from a rank $h\geq1$ in order to protect the metric against masking bias that occurs when outliers form small, dense clusters. Third, to accommodate complex geometries, pairwise distances are weighted using the inverse sample covariance matrix $\hat{\Sigma}^{-1}$. 

We emphasize that these modifications are not the primary theoretical contributions of this work; rather, they provide a well-conditioned outlier score to feed into our general thresholding procedure.

\begin{remark}
    \label{rem3:knn_ranks}
    The optimization of the neighborhood ranks $(h,l)$ is beyond the scope of this work. Throughout this manuscript, we adhere to standard asymptotic $k$-NN theory (Fix \& Hodges Jr., 1952 \cite{fix1952discriminatory}; Hall et al., 2008 \cite{hall2008choice}), setting $l$ such that $l \to \infty$ and $l/N \to 0$ as $N \to \infty$ (e.g., $l = \lfloor\sqrt{N}\rfloor$), which ensures estimator consistency (Cover \& Hart, 1967 \cite{cover1967nearest}). The lower rank is conventionally set to a small fraction of $l$ (e.g., $h = \lfloor 0.10 \cdot l \rfloor$).
\end{remark}

\subsection{Probabilistic False-Alarm Control}
\label{sec32:probabilistic_interpretation}
\noindent Setting the upper-bound of Eq.~\eqref{eq10:cantelli_base} equal to a target $\alpha \in (0,1]$ and solving for $c$ gives
\begin{equation}
    P\left(Y \geq \underbrace{\mu + \sigma \sqrt{\frac{1}{\alpha}-1}}_{T_\alpha}\right) \leq \alpha,
    \label{eq11:cantelli_derived}
\end{equation}

\noindent which can be interpreted as follows: \textit{under the null distribution of pseudo-isolations $Y$, the probability of a false positive cannot exceed $\alpha$}.\\

\noindent In practice, the theoretical parameters $(\mu, \sigma)$ are replaced by the running sample estimators $(\mu_m, \sigma_m)$ from Eq.~\eqref{eq2:running_stats}. This substitution is formally validated by the sample-based tolerance framework of Saw et al. (1984) \cite{saw1984chebyshev}:
\begin{equation}
    P\left(Y - \mu_m \ge c\sigma_m\right) \le \frac{1}{1 + c^2 \left(\frac{m}{m-1}\right)}.
    \label{eq12:guttman_bound}
\end{equation}
Since the finite-sample correction factor $\frac{m}{m-1} > 1$ for all $m > 1$, the empirical bound in Eq.~\eqref{eq12:guttman_bound} is slightly tighter than its population counterpart. This difference dissipates rapidly as $m$ grows (e.g., $\frac{m}{m-1} \approx 1.03$ at $m = 30$). Also, the conservative nature of this boundary is balanced by the Bessel-corrected sample standard deviation $\sigma_m$, which is a more generous estimator of scale when compared to its population counterpart.

\section{Inverse Inference on Significance Level \texorpdfstring{$\alpha^\star$}{alpha*}}
\label{sec4:inference}
The probabilistic framework described in Sections~\ref{sec21:outlier_threshold} and \ref{sec22:definitions} allows for an inverse inferential (local) approach that answers the following question: given an observation $i$, which is the minimum value $\alpha\in(0,1)$ such that $i$ is an outlier at that level? We will refer to said value as significance level $\alpha^\star_i$. More formally:
\begin{equation}
    \label{eq13:significance}
    \alpha_i^\star = \text{min}\{\alpha \in (0,1) : y_i\geq T_\alpha^\star\}.
\end{equation}
\noindent The evaluation of $\alpha^\star_i$ reveals a complex "outlier topology", which will be formalized in the following section.
\subsection{The Coalescence Property}
\label{sec41:coalescence}

We start by partitioning the (open) unit interval as follows:

\begin{equation}
    \label{eq14:partition}
    (0,1) = (0,\alpha_{0}) \cup \left(\bigcup_{q=0}^{Q-1} [\alpha_{q}, \alpha_{q+1})\right),
\end{equation}

\noindent where $\alpha_0 := \max\{\alpha \in (0,1) : y_{(N)}< T_{\alpha}^\star\}$ is the highest false-alarm rate at which no outlier is detected, and $\alpha_Q := 1$. This partition has three main properties:
\begin{enumerate}
    \item \textit{Quiescence}: The interval $(0,\alpha_0)$ is the only \textit{non-significant} outlierness interval, since no outlier below level $\alpha_0$ exists. In other words, the detection mechanism returns no outliers until $\alpha\geq\alpha_0$.
    \item \textit{Class Equivalence}: Let $N_{\alpha_q}=|\{i\in \{1,\dots,N\}: y_i \geq T^\star_{\alpha_q}\}|$ be the number of outliers at level $\alpha_q$. Then, $N_{\alpha}=N_{\alpha_q}$ $\forall \alpha \in [\alpha_q,\alpha_{q+1})$. In simpler terms, all levels in the generic interval $[\alpha_q,\alpha_{q+1})$ correspond to the same amount of outliers (holds trivially for $\alpha \in (0,\alpha_0)$).
    \item \textit{Coalescence}: Let $M_q = |\{i \in \{1,\dots,N\}: y_i \in [T^\star_{\alpha_q},T^\star_{\alpha_{q+1}})\}|$ be the number of outliers that need at least a significance level $\alpha_q$ to be detected. Then, $M_q\geq 1$. In other words, outliers move in "coalescence groups", and it is not necessarily possible to find $N-1$ thresholds that separate each unit from the others.
\end{enumerate}

\noindent The first property stands by the definition of $\alpha_0$. The second property follows by observing that $T^\star_{\alpha}$ (as a function of $\alpha$) is a continuous and strictly decreasing mapping of the interval $(0,1)$, while the set of ordered pseudo-isolations $\tilde{\mathbf{Y}}$ is finite. This means that the condition $y_i \geq T^\star_{\alpha_q}$ can only change for $\alpha_q$ in a finite set of significant levels $\{\alpha_0, \dots, \alpha_{Q-1}\}$, with $Q \in \{2,...,N\}$. Consequently, for any $\alpha$ within a sub-interval $(\alpha_q, \alpha_{q+1})$, the threshold $T_\alpha$ does not cross any observation in $\tilde{\mathbf{Y}}$, leaving the outlier count $N_{\alpha}=N_{\alpha_q}$ invariant. The third property is the core of the current discussion, and requires a rigorous proof. 

\begin{proof}
    \textit{(Coalescence)} Let $y_{(m)}$, $\mu_m$ and $\sigma_m$ be defined as in Section~\ref{sec21:outlier_threshold}. Assume, for the sake of contradiction, that $M_q=1$ for all $q$. This implies that for any $m \in \{2,\dots,N\}$, there \textbf{always} exists a level $\alpha\in(0,1)$ such that the unit with pseudo-isolation $y_{(m)}$ is considered anomalous, while the unit with pseudo isolation $y_{(m-1)}$ is not. Formally, this requires the following conditions to hold simultaneously:

\begin{equation}
    \label{eq15:proof1}
    \left\{
    \begin{aligned}
        &y_{(m)} \geq \mu_m + \sigma_m \sqrt{\alpha^{-1}-1} \\
        &y_{(m-1)} < \mu_{m-1} + \sigma_{m-1} \sqrt{\alpha^{-1}-1}\ .
    \end{aligned}
    \right.
\end{equation}
By isolating the shared constant $r=\sqrt{\alpha^{-1}-1}$, we have that:
\begin{equation}
    \label{eq16:proof2}
    L:=\frac{y_{(m)} - \mu_m}{\sigma_m} \leq r < \frac{y_{(m-1)} - \mu_{m-1}}{\sigma_{m-1}} :=U
\end{equation}
\noindent For the initial hypothesis to hold, the inequality $U > L$ must be satisfied for every rank $m$ across all valid ordered sequences of positive real numbers $\mathbf{\tilde{Y}}$. Consequently, exhibiting a single analytical instance where $U \le L$ is sufficient to prove the complementary hypothesis.\\

\noindent Consider the arbitrary set of ordered pseudo-isolations $\tilde{\mathbf{Y}} = (y_{(1)}, y_{(2)}, y_{(3)}) =(\sqrt{2}, e^\pi, 42)$ and suppose we want to separate $y_{(2)}=e^\pi$ and $y_{(3)}=42$ with a threshold $T_\alpha$. For the sub-sequence up to rank $m=2$, we obtain the running parameters $\mu_2 \approx 12.2774$ and $\sigma_2 \approx 15.3629$, which yield the upper bound:
\begin{equation}
    U = \frac{y_{(2)} - \mu_2}{\sigma_2} = \frac{e^{\pi} - 12.2774}{15.3629} \approx 0.7071
\end{equation}
On the other hand, evaluating the reference set up to rank $m=3$ yields $\mu_3 \approx 22.1850$ and $\sigma_3 \approx 20.3013$, determining the lower bound:
\begin{equation}
    L = \frac{y_{(3)} - \mu_3}{\sigma_3} = \frac{42 - 22.1850}{20.3013} \approx 0.9760
\end{equation}
Substituting these values into Eq.~\eqref{eq16:proof2} requires $0.9760 \le r < 0.7071$, which is absurd since the lower bound strictly exceeds the upper bound ($L_3 > U_3$). Because the parameter space for $r$ collapses into an empty set, no significance level $\alpha$ can isolate $y_{(3)}$ without simultaneously capturing $y_{(2)}$. This analytical counterexample disproves the starting assumption, establishing that there exists at least one coalescence group $q$ such that $M_q > 1$.
\end{proof}

\subsection{Packet-Boundary Search Problem}
\label{sec42:alpha_star}
\noindent In order to evaluate $\alpha_i^\star$, we define the mapping $\Phi: \mathbb{N} \to (0, 1)$ as:
\begin{equation}
\label{eq17:mapping}
    \Phi(m) = \left[\left(\frac{y_{(m)}-\mu_m}{\sigma_m}\right)^2+1\right]^{-1},
\end{equation}
\noindent where $y_{(m)},\ \mu_m$ and $\sigma_m$ are defined as in Section~\ref{sec21:outlier_threshold}. The significance level $\alpha^\star_i$ for a target unit $i$ with ordered pseudo-isolation $y_{(m)}$ is then:

\begin{equation}
    \label{eq18:significancelevel}
    \alpha_i^\star = \max_{j\geq m} \Phi(j)
\end{equation}

\noindent In other words, the solution of Eq.~\eqref{eq18:significancelevel} identifies the significance level of the initial element of the coalescence chain to which observation $i$ belongs. Clearly, if $j=m$ it follows that $y_i=y_{(m)}$ is exactly the initial element of its coalescence packet.

\section{Algorithms and computational complexity}
\label{sec5:algorithms} 
To simplify the illustration of our procedure, throughout this work we will set $\mathbf{\hat \Sigma} = \mathbf{I}$,  reducing the metric to the standard squared Euclidean distance. Nevertheless, the theoretical properties of the resulting architecture still hold under any arbitrary symmetric positive-definite covariance structure $\mathbf{\hat \Sigma}$. 
\subsection{Implementing ODK-means}
\label{sec51:odkm_implementation}
The ODK-means algorithm requires the data matrix $\mathbf{X}$ and the number of clusters $K$ as mandatory inputs. Users may also specify the trimming option (see Section~\ref{sec23:odkm}), the false-alarm rate $\alpha$ (defaulting to $\alpha=0.05$) and the neighborhood size parameters $h$ and $l$ (otherwise, they will be internally evaluated according to Remark~\ref{rem3:knn_ranks}).

\begin{enumerate}
    \item \textbf{Global Outlier Detection}: Identify \textit{External} and \textit{Internal} (not \textit{Cluster-specific}) outliers as described in Section~\ref{sec21:outlier_threshold}. Let $\mathbf{S}_0^\star$ denote the global null sample. 
    
    \item \textbf{Initialization}: Randomly select initial centroids from $\mathbf{S}_0^\star$.
    
    \item \textbf{Assignment}: Assign each data point to the nearest centroid to form the clusters $\mathbf{S}_1, \dots, \mathbf{S}_K$.
    
    \item \textbf{Local Outlier Detection}: Detect outliers within each cluster $\mathbf{S}_k$. This yields the cluster-specific null samples $\mathbf{S}_1^\star, \dots, \mathbf{S}_K^\star$. Assign weights to outliers according to Section~\ref{sec23:odkm} and update centroids.
    
    \item \textbf{Update and Convergence}: Repeat from Step (3) until the decrease in the objective function is less than a predetermined threshold $\epsilon$.
\end{enumerate}

\noindent Multiple restarts are recommended to mitigate the risk of converging to a local optimum, consistent with standard $K$-means. This ALS approach ensures a monotonic update of the objective function. Upon convergence, the algorithm returns the centroid matrix $\mathbf{\bar{X}}$, the assignment matrix $\mathbf{U}$ and the weights matrix $\mathbf{W}$.

\subsection{Computational Complexity}
\label{sec52:complexity}
The ODK-means algorithm is designed to shift the primary computational burden to a one-time pre-processing stage, ensuring that the iterative clustering phase remains as efficient as standard $K$-means. The complexity is decomposed as follows:

\begin{description}
    \item[Pre-processing Phase:] This phase is executed only once:
    \begin{itemize}
        \item \textit{Distance Matrix Calculation}: Evaluating the full $N \times N$ distance matrix $\mathbf{D}$ incurs a cost of $O(N^2 J)$, where $J$ is the dimensionality of the data.
        \item \textit{Isolation Extraction}: For each observation $i$, the pseudo-isolation value $y_i$ is extracted by finding the $h$-th to $l$-th smallest elements in the $i$-th row of $\mathbf{D}$. Using a selection algorithm (e.g., Quickselect), this takes $O(N)$ per row, totaling $O(N^2)$.
        \item \textit{Global Sorting}: The vector of isolations $\mathbf{Y}$ is sorted once at a cost of $O(N \log N)$.
    \end{itemize}
    The total cost of the pre-processing phase is therefore $O(N^2 J)$. 

    \item[Iterative Phase:] During the clustering process, the additional steps for outlier detection are highly optimized:
    \begin{itemize}
        \item \textit{Sub-sampling}: Since $\tilde{\mathbf{Y}}$ is sorted, extracting the values for a specific cluster $\mathbf{S}_k$ requires only $O(N_k)$, where $N_k$ is the cluster size.
        \item \textit{Threshold Optimization}: Evaluating ${}_k T_{\alpha}^\star$ involves calculating prefix sums in $O(N_k)$. The subsequent ALS search converges in a constant number of steps, each costing $O(\log N_k)$ via binary search.
    \end{itemize}
\end{description}

\noindent Summing over all $K$ clusters, the total per-iteration overhead is $\sum_{k=1}^K O(N_k) = O(N)$. Consequently, this phase maintains the same \textit{linear scalability per iteration} as standard $K$-means. While the initial investment is $O(N^2 J)$, the efficiency of the main loop makes it highly performant when the distance matrix can be stored in memory.

\subsection{Implementing the \texorpdfstring{$\alpha^\star$}{alpha*} search}
\label{sec53:alpha_star_algorithm}
The packet-boundary search procedure follows these steps:

\begin{enumerate}
    \item \textbf{Initial Estimation}: Define the initial reference set $\tilde{\mathbf{Y}}_m$ and compute the baseline "naive" significance level $\hat{\alpha}^\star_i = \Phi(m)$.
    \item \textbf{Outward Expansion}: Construct the candidate sequence of naive isolation levels for all remaining, more isolated ranks by evaluating $\alpha^\star_{j} = \Phi(j)$ for each $j \in \{m+1, \dots, N\}$.
    \item \textbf{Output}: Return the maximum value across the entire outward sequence.
\end{enumerate}

\noindent This procedure is designed for application to a single unit. A simple forward scan through the remaining pre-sorted subset of potential anomalies is computationally negligible when compared to the full clustering process.

\section{Simulations}
\label{sec6:simulations}

In this section, the performance of the ODK-means (ODKm) algorithm is evaluated against established R implementations of \textit{Trimmed $K$-means} (TKm) (Hennig 2025 \cite{trim25}) and the \textit{Local Outlier Factor} (LOF) (Hu et al. 2022 \cite{lof22}), thus covering both major distance-based and isolation-based outlier detection paradigms. \textit{(A detailed description of the simulation design and spatial filtering process is provided in Section S3 of the Supplementary Material.)}

\begin{figure}[H]
    \centering
    \includegraphics[width = 0.85\textwidth]{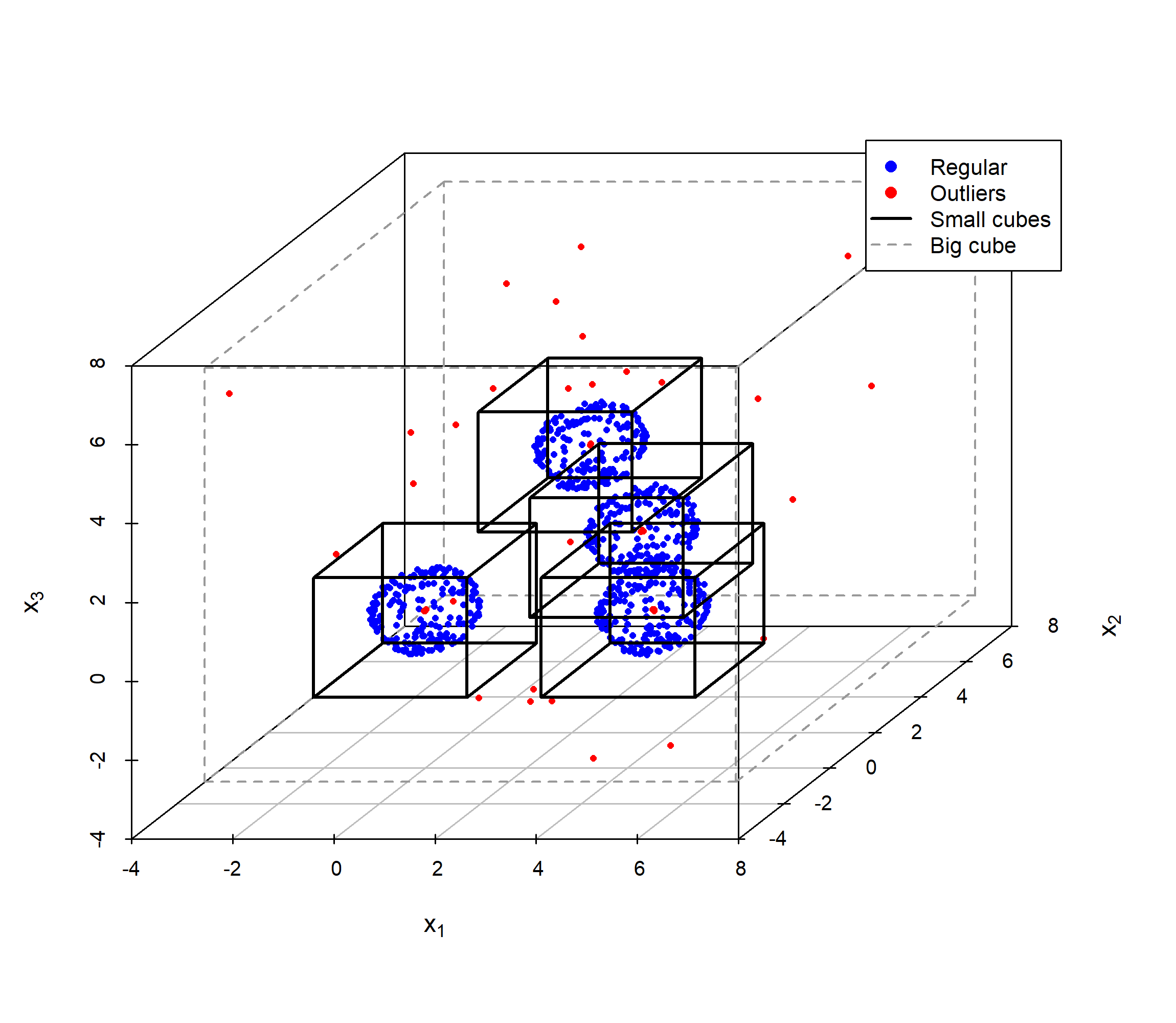}
    \caption{Illustration of a synthetic dataset realization in a three-dimensional feature space ($N_k=200, r=1, s=4.5, \sigma^2=0.02, p=0.05$). Regular observations and anomalies are shown in blue and red, respectively. The inner cubes around each cluster and the outer bounding box define the spatial boundaries used to control the generation of internal versus external outliers according to the definitions in Section~\ref{sec22:definitions}.}
    \label{fig2:dataset_generation}
\end{figure}

\noindent To ensure high variability across testing environments, synthetic datasets are generated by tuning five structural parameters: cluster size ($N_k$), cluster radius ($r$), inter-cluster distance ($s$), noise variance ($\sigma^2$) and contamination rate ($p$). Figure~\ref{fig2:dataset_generation} illustrates the geometric construction process for a sample dataset. Given the inherent class imbalance between regular units and anomalies, the $F_1$ score provides a robust metric that balances precision and sensitivity without being inflated by true negatives. Since both Trimmed $K$-means and the LOF need a proportion $\hat{p}$ of items to flag (information unavailable for real-data), tests will be repeated with $\hat{p}=p$, $\hat{p}=0.75p$ and $\hat{p}=1.25p$ to check robustness. On the other hand, ODK-means will always run with a False-alarm rate $\alpha=0.05$. For each parameters combination, algorithms are evaluated across 10 simulated datasets, totaling 1050 runs ($>20000$ considering 20 random starts per run). Performance is reported as the average $F_1$ score across replicates.

\begin{remark}
    \label{rem4:kmeans_limitations}
    The proposed outlier detection mechanism is not tied to $K$-means and can be incorporated into various clustering procedures. Consequently, ODK-means inherits all limitations of standard $K$-means, except those related to the presence of outliers. For this reason, our simulation study focuses exclusively on settings where clusters are approximately $J$-spheres. Testing non-spherical clusters would add little value, since it is already well established that $K$-means performs poorly in such situations. In practice, the detection module can be combined with clustering algorithms that are robust to more general cluster geometries (e.g., Gaussian-mixture models with flexible covariance structures).
\end{remark}

\begin{remark}
    \label{rem5:lof_thresholding}
    To accommodate the evaluation of the LOF algorithm, which outputs a continuous anomaly score rather than a discrete classification, a uniform thresholding procedure is adopted. For a given input proportion \( \hat{p} \in (0, 1) \), the top $\lfloor N\hat{p}\rfloor$ instances with the highest LOF scores within each dataset are flagged as anomalous. This dynamic thresholding approach follows the original framework and avoids inefficient, per-dataset parameter tuning. Furthermore, since LOF performs outlier detection without clustering, it is paired with standard $K$-means to obtain the final partition. 
\end{remark}

\subsection{First scenario: varying true outlier rate \texorpdfstring{$p$}{p}}
\label{sec61:variable_p}
Table~\ref{tab1:variable_p} shows the comprehensive simulation result. 

\begin{table}[!ht]
    \centering
    \caption{Comparison of average F1 scores for ODK-means ($\alpha = 0.05$) and competing procedures for $\hat{p}=p$ and $\hat{p}\neq p$ with variable $p$. Fixed data generation parameters: $N_k=300$, $r=3$, $s=8$, $\sigma^2=0.1$}
    \label{tab1:variable_p}
    \small 
    \begin{tabular}{lcccccccc}
        \toprule
        & & \multicolumn{2}{c}{$\hat{p} = p$} & & \multicolumn{4}{c}{$\hat{p} \neq p$} \\
        \cmidrule(lr){3-4} \cmidrule(lr){6-9}
        $p$ & ODK-means & TKm & LOF & & \multicolumn{2}{c}{TKm} & \multicolumn{2}{c}{LOF} \\
        & $\alpha = 0.05$ & & & & $0.75p$ & $1.25p$ & $0.75p$ & $1.25p$ \\
        \midrule
        $0.0025$ & 1.000 & 1.000 & 0.800 & & 1.000 & 0.857 & 1.000 & 1.000 \\
        $0.02$   & 1.000 & 0.700 & 0.781 & & 0.772 & 0.641 & 0.592 & 0.625 \\
        $0.05$   & 1.000 & 0.700 & 0.410 & & 0.777 & 0.636 & 0.306 & 0.277 \\
        \bottomrule
    \end{tabular}
\end{table}

\noindent Both TKm and the LOF are highly sensitive to the accuracy of $\hat{p}$. The perfect score of TKm at the lowest contamination rate ($p=0.0025$) when $\hat{p}=p$ occurs due to the absence of cluster-specific outliers. By dataset construction, given four clusters of size $N_k=300\ (k=1,...,4)$, there will be a total of $0.0025\cdot4\cdot300=3$ anomalies, $\lfloor{0.3\cdot3}\rfloor=\lfloor0.9\rfloor=0$ of which will be cluster-specific. This scenario is entirely within the algorithm's design strengths. The capabilities of TKm when $\hat{p}$ is correct are reflected by an F1 score of $0.7$. In fact, since this procedure is not able to detect cluster-specific outliers (by construction), all other outliers have been identified in every single run. In the presence of cluster-specific outliers, underestimating $p$ results in a higher performance, since no false positives are flagged. To be more precise, given $p_1$ and $p_2$ the proportion of cluster-specific and non-cluster-specific outliers ($p=p_1+p_2$) respectively, TKm will return about $(\hat{p}-p_2)N$ false positives if $\hat{p}>p_2$ and $(p_2-\hat{p})N$ false negatives otherwise. This observation is reflected in the performance of TKm for $\hat{p}=1.25p$, which is by far the worst among all scenarios. In contrast, LOF's performance declines substantially as the true contamination rate $p$ increases. A higher density of outliers fills the naturally sparse regions between clusters, making the overall data distribution appear more uniform and thus obscuring the local density deviations that LOF relies upon. Finally, ODKm demonstrates perfect outlier detection across all tested contamination rates, successfully identifying all kinds of outliers.

\subsection{Second scenario: varying data dispersion \texorpdfstring{$\sigma^2$}{sigma2}}
\label{sec62:variable_sigma}
Table~\ref{tab2:variable_sigma} highlights TKm's robustness against increases in $\sigma^2$. This is an expected outcome, since a homogeneous increase in data dispersion simply scales distances. As already observed, the best performance of TKm is obtained when $\hat{p}<p$. On the other hand, the LOF exhibits a weaker, though still present, robustness against increments of $\sigma^2$. In fact, the homogeneous increase in dispersion uniformly shifts the LOF towards higher values. While ODKm maintains superior overall performance, higher dispersion causes a slight decline in its score. This minor drop is mainly an artifact of the data generation mechanism rather than a theoretical limitation. Specifically, as $\sigma^2$ increases, the boundaries between regular observations and "true" cluster-specific anomalies overlap.

\begin{table}[!ht]
    \centering
    \caption{Comparison of average F1 scores for ODK-means ($\alpha = 0.05$) and competing procedures for $\hat{p}=p$ and $\hat{p}\neq p$ with variable $\sigma^2$. Fixed data generation parameters: $N_k=300$, $r=3$, $s=8$, $p=0.02$}
    \label{tab2:variable_sigma}
    \small 
    \begin{tabular}{lcccccccc}
        \toprule
        & & \multicolumn{2}{c}{$\hat{p} = p$} & & \multicolumn{4}{c}{$\hat{p} \neq p$} \\
        \cmidrule(lr){3-4} \cmidrule(lr){6-9}
        $\sigma^2$ & ODK-means & TKm & LOF & & \multicolumn{2}{c}{TKm} & \multicolumn{2}{c}{LOF} \\
        & $\alpha = 0.05$ & & & & $0.75p$ & $1.25p$ & $0.75p$ & $1.25p$ \\
        \midrule
        $0.02$ & 1.000 & 0.700 & 0.625 & & 0.772 & 0.607 & 0.725 & 0.610 \\
        $0.2$  & 0.998 & 0.700 & 0.612 & & 0.772 & 0.607 & 0.601 & 0.581 \\
        $0.5$  & 0.995 & 0.700 & 0.608 & & 0.772 & 0.607 & 0.594 & 0.552 \\
        \bottomrule
    \end{tabular}
\end{table}

\noindent All algorithms proved robust to variations in the remaining parameters, and these secondary results are omitted for brevity. Modifying cluster size ($N_k$) produces an effect equivalent to varying the true outlier rate ($p$). Similarly, increasing the cluster radius ($r$) is equivalent to decreasing the inter-cluster distance ($s$).

\section{Methodological tutorial}
\label{sec7:tutorial}
To complement the theoretical results and simulation study, this section provides an in-depth illustration of how to apply ODK-means and utilize its core inferential capabilities. Since the performance of $K$-means is already well established, testing our proposal on large-scale real-world datasets would primarily evaluate the clustering engine rather than the anomaly detection mechanism, especially given the lack of ground-truth outlier labels. Instead, we use low-dimensional benchmark datasets that offer a clear visual environment to inspect the statistical validity of our framework. The Old Faithful Geyser dataset \cite{azzalini1990look} serves as an ideal testbed for this tutorial: its two features naturally form two primary groups, the low-density region contains ambiguous "bridging" elements and its bivariate structure allows for intuitive visual verification.

To ensure a fair comparison of outlier detection techniques, all required parameters were methodically calibrated. ODK-means was executed with a nominal false-alarm rate $\alpha = 0.05$ to set a baseline. As expected, the number of detected outliers ($5$) does not correspond to a fixed 5\% of the data ($\simeq 14$). For Trimmed $K$-means, the diagnostic routines in the \texttt{TClust} package \cite{tclust} confirmed $K = 2$ as optimal, but yielded no clear elbow for the trimming proportion $\hat{p}$; thus, we set $\hat{p} = 0.018$ to detect 5 outliers for direct alignment with ODK-means. Finally, as the \texttt{Rlof} package lacks an automated threshold selection tool, we evaluated two distinct cutoffs: $\text{LOF} = 1.70$ to detect 5 outliers and $\text{LOF} = 1.38$, corresponding to an elbow in the score distribution. \textit{(If interested in the diagnostic plots used to tune the parameters above, the reader is invited to check Section S4 of the Supplementary Material.)}

\subsection{Adaptive outlier detection}
\label{sec71:comparison}

\begin{figure}[H]
    \centering
    \includegraphics[width=\linewidth]{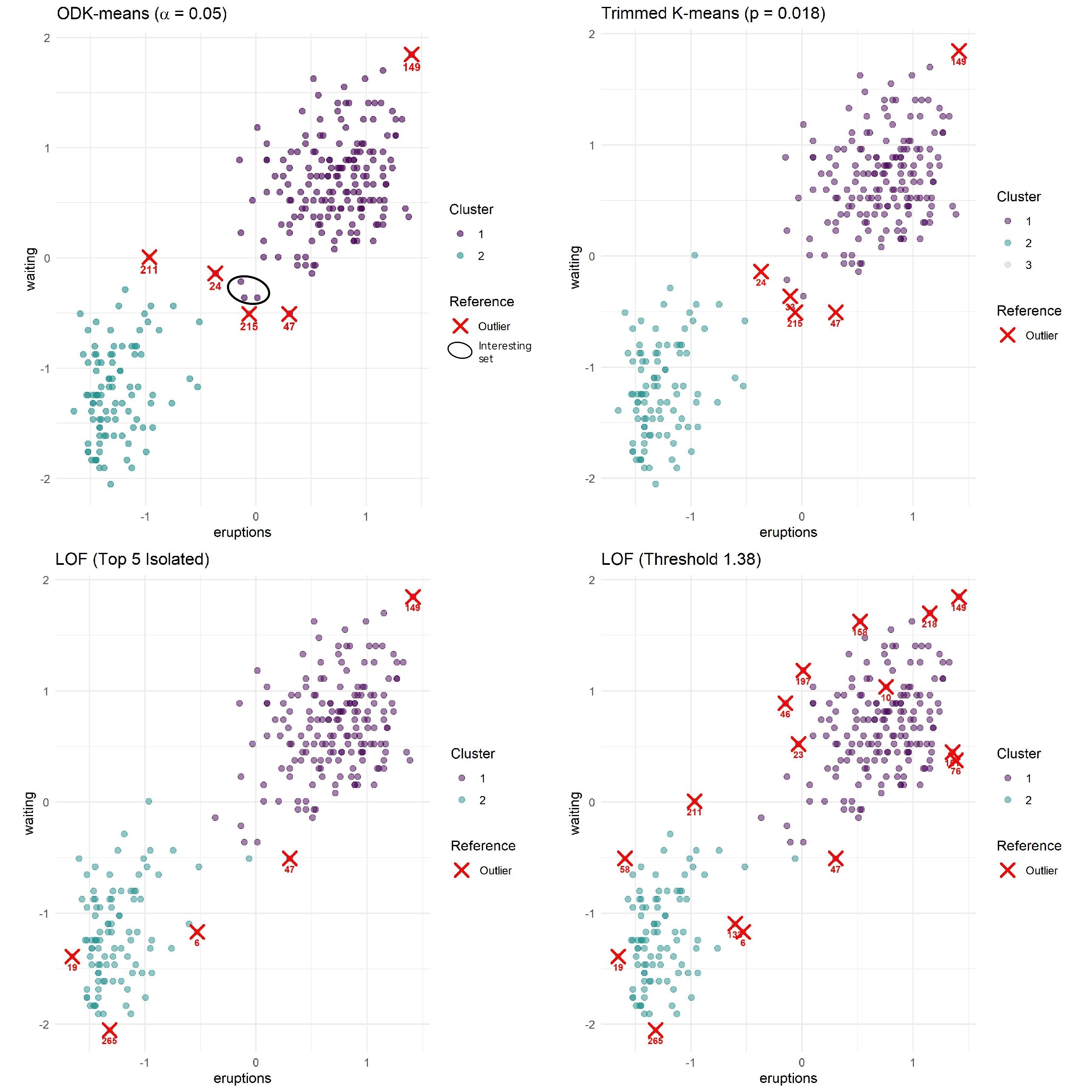}
    \caption{ODK-means at $\alpha=0.05$ (top-left); Trimmed $K$-means at $\hat{p}=0.018$ (top-right); LOF$=1.7$ + $K$-means (bottom-left); LOF$=1.38$ + $K$-means (bottom-right). Crossed units represent detected outliers (Note: Trimmed $K$-means places outliers in an additional cluster).}
    \label{fig3:geyser_comparison}
\end{figure}

Figure~\ref{fig3:geyser_comparison} illustrates how ODK-means, Trimmed $K$-means and LOF operate in fundamentally different ways. Noticeably, even though ODK-means relies on an isolation-based outlier score similar to LOF's, its overall detection profile behaves much more like Trimmed $K$-means. Both LOF configurations primarily target observations on the outer edges of individual clusters, while ODK-means and Trimmed $K$-means focus on the "more ambiguous" bridging elements between clusters. In that specific area, since average data dispersion is high, each item's LOF scores below the target threshold. As a result, LOF acts more like a "boundary smoother" rather than an anomaly detector.

On the other hand, ODK-means and Trimmed $K$-means show consensus on four detected items out of five: elements 24, 215, 47 (all falling in our "internal outlier" categorization) and 149 (which is instead an "external outlier"). Since real-world data lacks ground-truth outlier labels, we cannot declare a single algorithm universally correct. However, this high degree of alignment between two totally different frameworks mutually reinforces the validity of the identified subset. Nevertheless, the primary advantage of ODK-means over Trimmed $K$-means (and LOF) lies in its inferential framework. Unlike competing procedures, ODK-means returns anomalies equipped with a formal statistical interpretation. Specifically, under the assumed null distribution of pseudo-isolations, each detected unit (24, 47, 149, 211 and 215) has a probability $\alpha \leq 0.05$ of being a false alarm. ODK-means thus replaces heuristic trimming with a formal, testable statistical property.

\subsection{Inferential and topological features}
\label{sec72:inference_application}

In the low-density region between the two clusters in Figure~\ref{fig3:geyser_comparison} (top-left), three observations have been circled: they are (from left to right) units 174, 33 and 165. Even though sharing similar features with some crossed units, they were not flagged as outliers. For this reason, one might be interested in determining the false-alarm rate required for ODK-means to detect them. The procedure in Section~\ref{sec42:alpha_star} returns $\alpha^\star_{\{174,33,165\}}=\alpha^\star_{33} \simeq 0.0776$.

An additional possible application of $\alpha^\star$ is the evaluation of the \textit{actual} false-alarm rate within a set of anomalies. The least isolated outlier (in this case, unit 24) necessarily possesses a significance level $\alpha^\star_{24} \leq 0.05$. By applying the iterative search procedure, we find that $\alpha^\star_{24} \simeq 0.04694$. 

\begin{figure}[H]
    \centering
    \includegraphics[width=\linewidth]{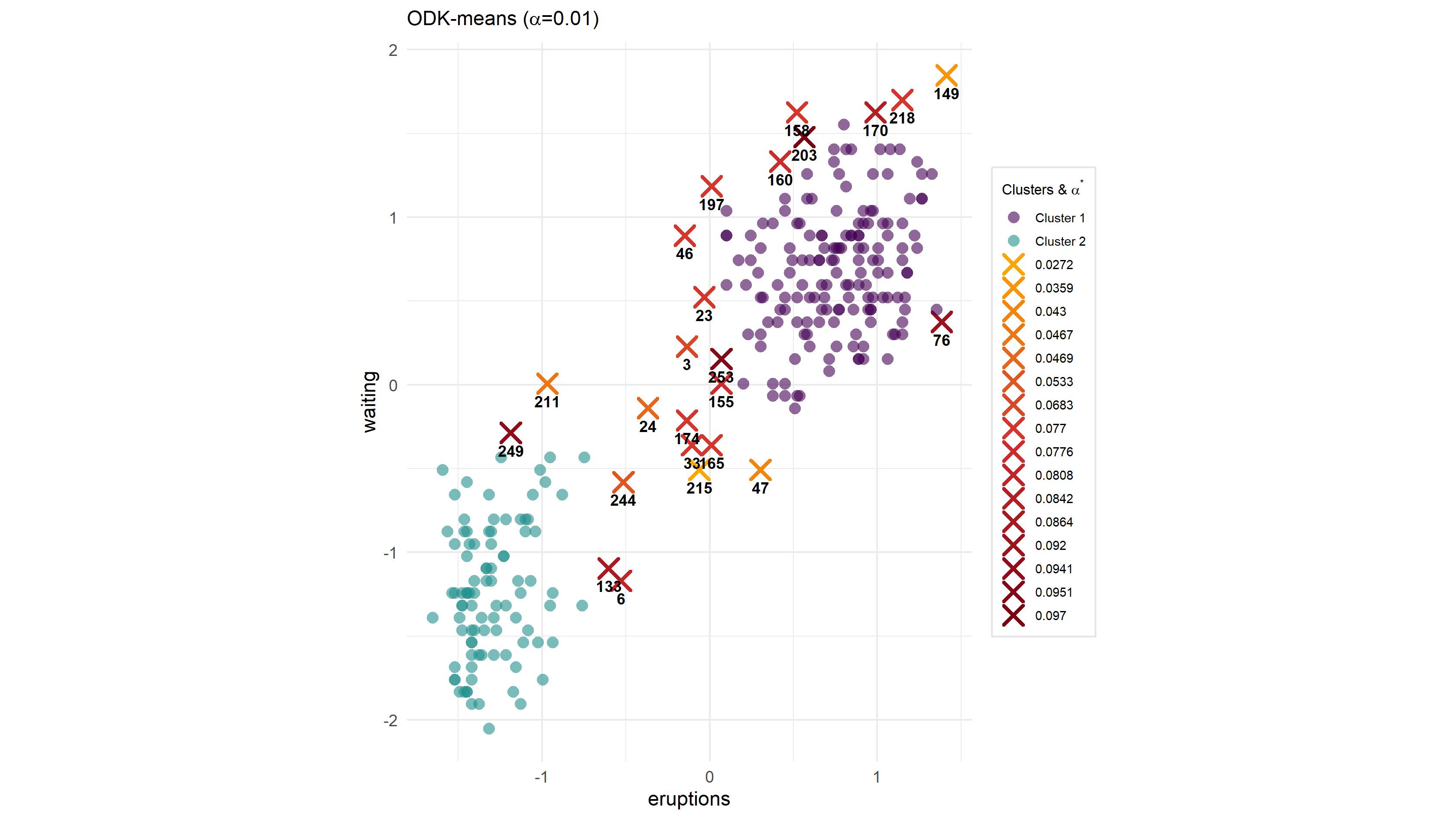}
    \caption{ODK-means results at $\alpha=0.1$. Outlier units sharing membership in the same coalescence packet are marked with same cross color.}
    \label{fig4:coalescence_geyser}
\end{figure}

\noindent Figure~\ref{fig4:coalescence_geyser} illustrates the results of ODK-means with a false-alarm rate $\alpha = 0.1$; its inspection is not intended to differentiate individual packet boundaries, but to visually validate all the structural properties derived in Section~\ref{sec41:coalescence}. Specifically: no anomalies are flagged with $\alpha \in(0, 0.0272)$, with unit $215$ emerging as the most extreme outlier that initiates the detection chain (\textit{Quiescence}); the topology remains invariant outside the specific packet thresholds shown in the legend (\textit{Class Equivalence}); the system detects a total of 24 outliers with only 16 unique significance levels, showing that multiple distinct observations collapse into shared, simultaneous detection gates (\textit{Coalescence}).

\subsection{Bonus scenario: No outliers}
\label{sec73:iris}

\begin{figure}[H]
    \centering
    \includegraphics[width=\linewidth]{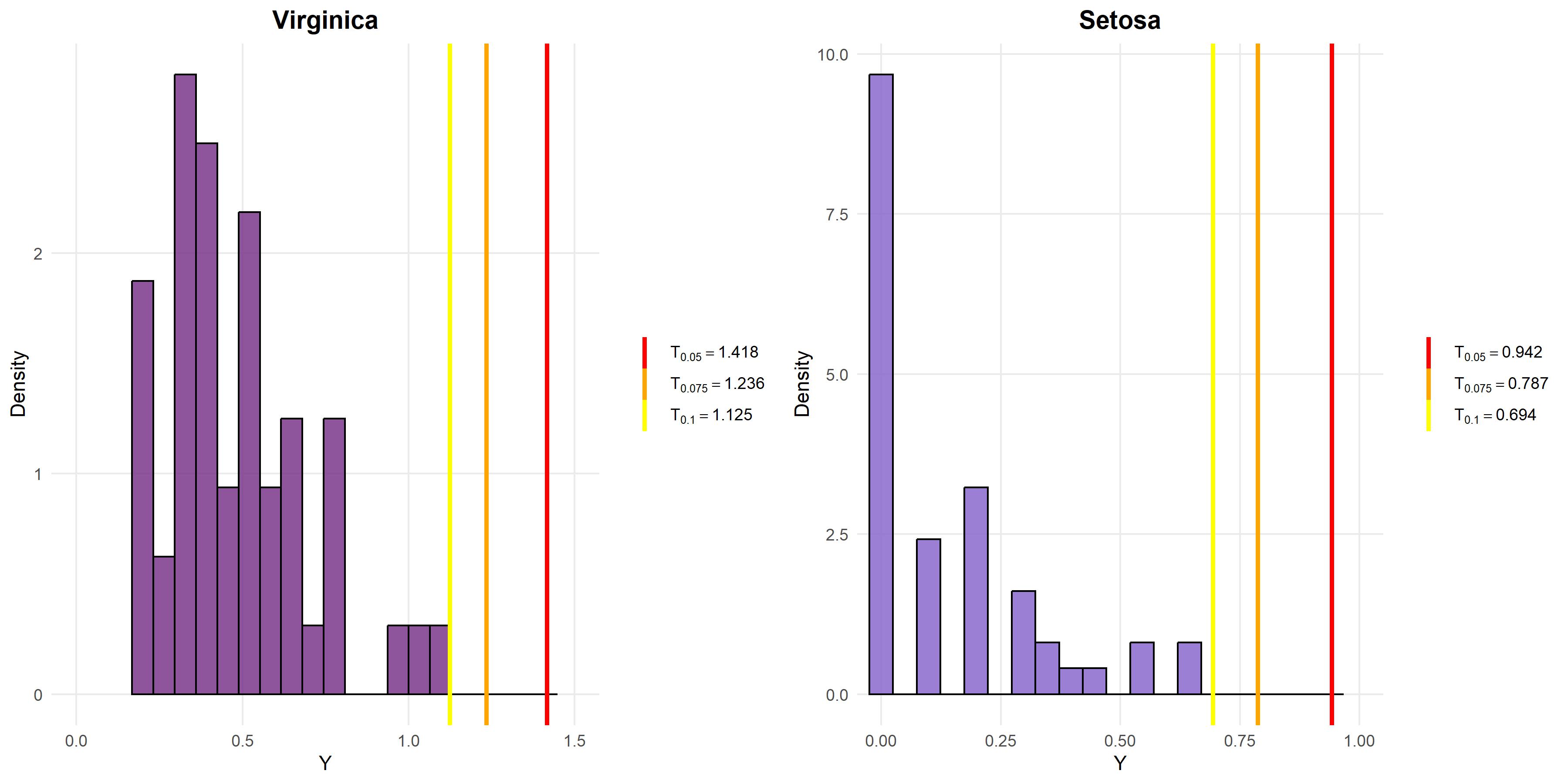}
    \caption{ODK-means applied to Setosa and Virginica species from the Iris dataset ($\alpha = \{0.05,0.075,0.10$\}). All observations remain categorized as regular.}
    \label{fig5:iris_no_outliers}
\end{figure}

To empirically show how ODK-means reduces to standard $K$-means when no anomalies are detected, we isolate the Setosa and Virginica species from the Iris dataset \cite{Iris} using their Petal Length and Petal Width measurements to form two anomaly-free clusters. As shown in Figure~\ref{fig5:iris_no_outliers}, even when raising the false-alarm rate to $\alpha = 0.10$ (a value unusually high for practical applications), the outlier thresholds consistently remain beyond the main body of both pseudo-isolation distributions. This behavior is particularly remarkable given the inherent modularity and overlap within the Iris dataset. Since the trimming step is bypassed, equal weights are assigned to all observations just like in $K$-means. To the best of our knowledge, this feature remains unique in the literature. Consider the case of Trimmed $K$-means: while one could theoretically set the trimming proportion to zero ($\hat{p}=0$), this would create a logical paradox. The prior knowledge of no outliers would in fact suggest the use of the more efficient standard $K$-means algorithm. 

\section{Discussion}
\label{sec8:discussion}

This paper introduced a novel, distribution-agnostic and probabilistic definition of outlier to develop a computationally efficient detection mechanism. By integrating this framework directly into $K$-means, we proposed the ODK-means model, whose capabilities are demonstrated in numerous applications. 

It is worth emphasizing that the proposed framework is not limited to the simple settings shown in the methodological tutorial. Its modular logic extends naturally to more complex geometries and high-dimensional settings. For example, non-spherical or overlapping clusters can be handled by incorporating our detection mechanism into Gaussian Mixture Models (GMMs), as discussed in Remark~\ref{rem4:kmeans_limitations}. To address the curse of dimensionality, the same logic can be applied to subspace clustering techniques, such as Reduced $K$-means \cite{de1994k} or Factorial $K$-means \cite{vichi2001factorial}. Finally, the core isolation score itself can be tailored to handle specific structural constraints, such as multi-way data arrays or disjoint component constraints, as treated by XT3Clus \cite{bottazzi2026three}. Choosing $K$-means as our baseline clustering algorithm provided a clear and intuitive foundation, allowing us to showcase the visual mechanics of the proposed framework with minimal complexity. Furthermore, this choice ensures a direct and fair comparison against the most established outlier detection procedures in the literature.

Since our proposal has been developed from a purely methodological perspective, the full range of applicability of its theoretical properties may be partially immediate. While advantages like dynamic outlier selection or a clear false-alarm interpretation are easy to appreciate, the related inferential insights offer practical value in applied settings. For instance, in lab environments or controlled baseline trials, researchers often face ambiguous boundary observations. Rather than relying on rigid, arbitrary cutoffs, evaluating the exact false-alarm rate $\alpha^\star$ required to flag those specific subjects allows scientists to assess the exact risk/tolerance needed to consider them anomalous. To give another example, when domain experts work with confirmed historical outliers, the evaluation of their $\alpha^\star$ provides a theoretically grounded benchmark to calibrate false-alarm budgets for future studies.

We conclude this work by pointing to the main direction for future research: developing a fully parametric extension of this framework to derive a closed-form distribution for the pseudo-isolation score. Achieving an analytical expression for this distribution will enable formal, exact hypothesis testing for individual outliers, providing an even stronger inferential foundation for anomaly detection.

\section*{Code Availability}
\begin{itemize}
    \item \textbf{Software and Implementation:} The R code supporting the findings of this study is available in the repository at \url{https://github.com/IannaccioStat/Outliers1}.
\end{itemize}
\newpage
\appendix
\section*{Supplementary material}

\section*{S1\hspace{8pt}A deeper literature review}
\label{sec:intro}
In Section 1, we highlighted how both "classic" and "contemporary" definitions of outliers can hardly be translated into detection algorithms. Here are a few examples, organized in chronological order: Anscombe (1960) \cite{anscombe1960rejection} points out that outliers are "far from the central mass of data", where the mass is defined as the smallest polytope that contains regular data. Tukey (1962) \cite{tukey1962future} emphasizes that outliers are data points that fall outside the expected range, particularly identified through techniques like the box plot. Hawkins (1980) \cite{hawkins1980identification} characterizes them as "observations that deviate so much from other observations as to arouse suspicions that they were generated by a different mechanism". Barnett et al. (1994) \cite{barnett1994outliers} describe outliers as "observations which appear to be inconsistent with the remainder of that set of data". Chandola et al. (2009) \cite{chandola2009anomaly} describe them as items that do not conform to a defined notion of normal behavior. Aggarwal (2016) \cite{aggarwal2016introduction} refers to outliers as the result of an unusual behavior of one or more generating processes.\\

\noindent A secondary concern of the outlier literature is the categorization of anomalies: According to Rousseeuw \& Leroy (2003) \cite{rousseeuw2003robust}, outliers can be categorized by data dimensionality: \textit{univariate outliers} are detectable by examining a single variable, whereas \textit{multivariate outliers} become apparent only when analyzing the joint behavior of multiple variables. Hodge \& Austin (2004) \cite{hodge2004survey} distinguish \textit{local} and \textit{global}  outliers. The former may not be unusual when viewed in the context of the entire dataset, but are anomalous when viewed in the context of their local neighborhood. The latter are anomalous with respect to the entire dataset. This distinction is especially relevant in datasets with spatial, temporal or manifold structures. Chandola et al. (2009) \cite{chandola2009anomaly} consider the \textit{collective outliers}, identified as groups of units that exhibit unusual behavior when studied together, rather than individually. This classification is particularly useful for detecting coordinated anomalies such as fraud in financial transactions that may collectively indicate an anomalous pattern. The work by Kriegel et al. (2012) \cite{kriegel2012outlier} introduces \textit{subspace outliers}, which are anomalies detectable only in specific feature subspaces, addressing challenges in high-dimensional data. Ahmed et al. (2016) \cite{ahmed2016survey} consider the \textit{contextual outliers}, which are observations that appear anomalous within a specific context or environment. As an example: a temperature of 37°C may be normal in summer but considered an outlier in winter. Zhao \& Yang (2019) \cite{zhao2019robust} extend this concept to spatio-temporal contexts, proposing adaptive thresholds for dynamic environments. The categorization provided in our paper (external, internal and cluster-specific outliers) is inspired by recent work in topological data analysis (TDA), where internal outliers correspond to “holes” or low-density regions within the null sample $\mathbf{S}^\star$ (Chazal 2017 \cite{chazal2017high}).\\

\noindent Since Trimmed $K$-means and the Local Outlier Factor represent the canonical baseline paradigms of distance-based and density-based robust partitioning, they are the primary benchmarks for our empirical evaluations. However, the robust clustering literature offers a rich spectrum of alternative proposals. The $K$-Medoids (PAM: Partitioning Around Medoids - Kaufman \& Rousseeuw 2009 \cite{kaufman2009finding}) algorithm utilizes medoids (observed units) instead of centroids to represent cluster, showing less sensitivity to outliers than traditional $K$-Means. Spherical $K$-Means (Dhillon 2001 \cite{dhillon2001co}) is a variant of $K$-Means that employs cosine similarity instead of Euclidean distance, making it suited for high-dimensional data such as text or document clustering. Outliers can be identified based on their angular distance from cluster centers, which tends to be larger for anomalies. Robust $K$-Means (Xu \& Wunsch 2005 \cite{xu2005survey}) modifies the $K$-Means algorithm by minimizing the median distance rather than the mean distance to adjust centroids. Other simple solutions to identify outliers
before clustering include Yu et al. (2016) \cite{yu2016outlier}: the OEDP $K$-means algorithm performs data preprocessing, avoiding the initial random centroids to be outlying datapoints. At first, select and remove outliers based on a density threshold; then, assign centroids and perform standard $K$-means on the whole dataset. In our opinion, the effects of such a useful tool should not be limited to the earliest stages of the algorithm, mainly for two reasons: firstly, cluster-specific outliers would not be detected; secondly, the information brought by anomalies  might get lost if they are only reintroduced in the analysis when the clusters are formed. More advanced clustering frameworks are also available, such as Non-Exhaustive, Overlapping $K$-Means (NEO-$K$-Means, Whang et al. 2015 \cite{whang2015non}), which is a reformulation of $K$-means that allows points to belong to multiple clusters or none. Points that do not belong significantly to any cluster may be classified as outliers. Finally, methods like $K$-Means with outlier removal (KMOR - Gan \& Ng 2017 \cite{gan2017k}) provide data clustering and outlier detection simultaneously by introducing an additional “cluster” to the predefined $K$ to hold all outliers. \\

\noindent Outlier detection can also be a byproduct of procedures with unrelated goals. A few examples include fuzzy algorithms such as Fuzzy C-Means (Dunn 1973 \cite{dunn1973fuzzy}; Bezdek et al. 1984 \cite{bezdek1984fcm}) and Density-based methods like DBSCAN (Ester et al. 1996 \cite{ester1996density}). The former allows points to belong to multiple clusters with varying degrees of membership. Here, outliers can be identified by their low membership values across all clusters, indicating that they do not fit well within any specific group (similarly to NEO-$K$-means). The latter is effective for linking outliers to noise. However, its limitations in high-dimensional spaces have spurred variants like HDBSCAN (Campello et al. 2013 \cite{campello2013density}), which automatically optimizes density thresholds. Rousseeuw et al. (2018) \cite{rousseeuw2018measure} provides a comprehensive summary of the topic and additional procedures for anomaly detection.

\section*{S2\hspace{8pt}Case-wise v. Cell-wise Outliers}
\label{sec:64}
\begin{figure}[H]
    \centering
    \includegraphics[width=\linewidth]{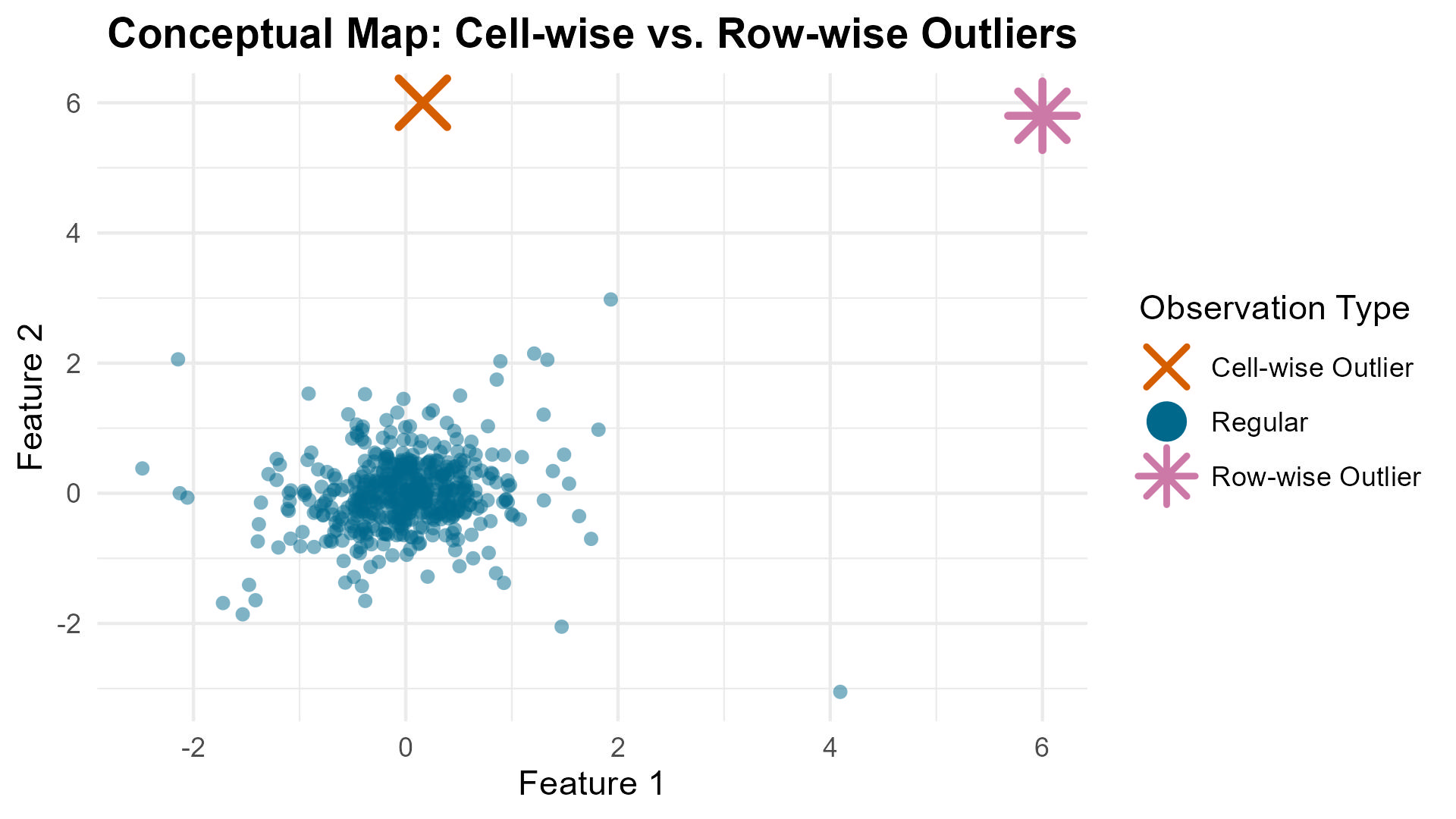}
    \caption{Illustration of cell-wise and case-wise outliers in 2 dimensions.}
    \label{figs1:case/cell}
\end{figure}

In recent years, the distinction between case-wise (Figure \ref{figs1:case/cell}, pink star) and cell-wise (Figure \ref{figs1:case/cell}, red cross) outliers has become a prominent topic in robust statistics. The former refers to units that show an overall deviation from the main mass of the data, the latter to units that exhibit abnormal behavior in only a small subset of features. As Alqallaf et al. (2009) \cite{alqallaf2009} highlight, cell-wise anomalies are especially challenging in high-dimensional settings: as features multiply, the probability that an observation contains at least one anomalous cell approaches 1, causing traditional case-wise routines to trim large portions of the data.

\noindent Since our methodology relies on a probabilistic bound, contamination is detected regardless of its underlying mechanism. While a full cell-wise diagnostic is beyond the scope of our current proposal, our logic extends intuitively to this setting. In fact, a researcher can easily diagnose flagged units $\mathbf{x}_i \in \mathbf{S}\setminus \mathbf{S}^\star$ by applying the proposed framework to the univariate pseudo-isolation of each component $x_{ij}$. Flagging components that exceed a feature-specific threshold $T_{\alpha}^{j\star}$ allows users to distinguish cell-wise from case-wise outliers based on a predefined feature cutoff $F < J$. If interested, the reader is redirected to established approaches such as Rousseeuw \& Bossche (2018) \cite{rousseeuw2018detecting}, who leverage correlation structures to predict expected cell ranges, or Raymaekers \& Rousseeuw (2024) \cite{raymaekers2024challenges}, who treat anomalous cells as missing values. 

\section*{S3\hspace{8pt}Data generation mechanism}

The simulation study in Section 6 relies on a generation mechanism that has been developed to consistently test outlier detection capabilities. The algorithm operates as follows. Four datasets of size $N_k$ $(k=1,...,4)$, each consisting of points uniformly distributed on the surface of a $J$-sphere with radius $r$, are generated. These four datasets are then positioned at the vertices of a tetrahedron with side length $s$, representing four clusters. Then, each observation is perturbed by a noise vector $\boldsymbol{\eta}_i = (\eta_{i1}, \dots, \eta_{iJ})^\prime$, where $\eta_{ic} \sim N(0, \sigma^2)$ for $c = 1, \dots, J$. Following this, a random sample of $\lfloor Np \rfloor$ data points is removed from the dataset, where $p \in (0,1)$. These removed points are replaced with artificial anomalies, consisting of 30\% cluster-specific and 70\% non-cluster-specific outliers. However, controlling "ground truth" anomalies in synthetic experiments presents a well-known challenge: purely random noise generation can easily spawn points near the core data mass, causing them to fall into high-density regions where they no longer behave as genuine outliers. To prevent this contamination and guarantee a clear separation of densities, we apply a strict spatial filtering procedure. Candidate anomalies are sampled within a bounding box enclosing the entire layout (side length $s + 3r$), but any candidate that falls inside a safety cube (side length $2.4r$) centered on any cluster is immediately rejected and resampled until the required number of anomalies is obtained. By adjusting the parameter set $(N_k, r, s, \sigma^2, p)$, we can simulate a wide range of experimental conditions. The cluster size $N_k$ tests the algorithm's scalability in large-sample environments. Adjusting the sphere radius $r$ alongside the cluster separation $s$ directly regulates the degree of cluster overlap, allowing us to evaluate performance under varying levels of spatial ambiguity. Meanwhile, modifying the noise variance $\sigma^2$ controls the dispersion around the sphere surfaces, which creates low-density transition zones along cluster boundaries that challenge the distinction between regular edge points and true anomalies. Finally, adjusting $p$ alters the overall contamination rate to assess detection robustness under light-to-heavy anomaly loads.

\section*{S4\hspace{8pt}Competing procedures tuning}
\begin{figure}[H]
    \centering
    \includegraphics[width=0.7\linewidth]{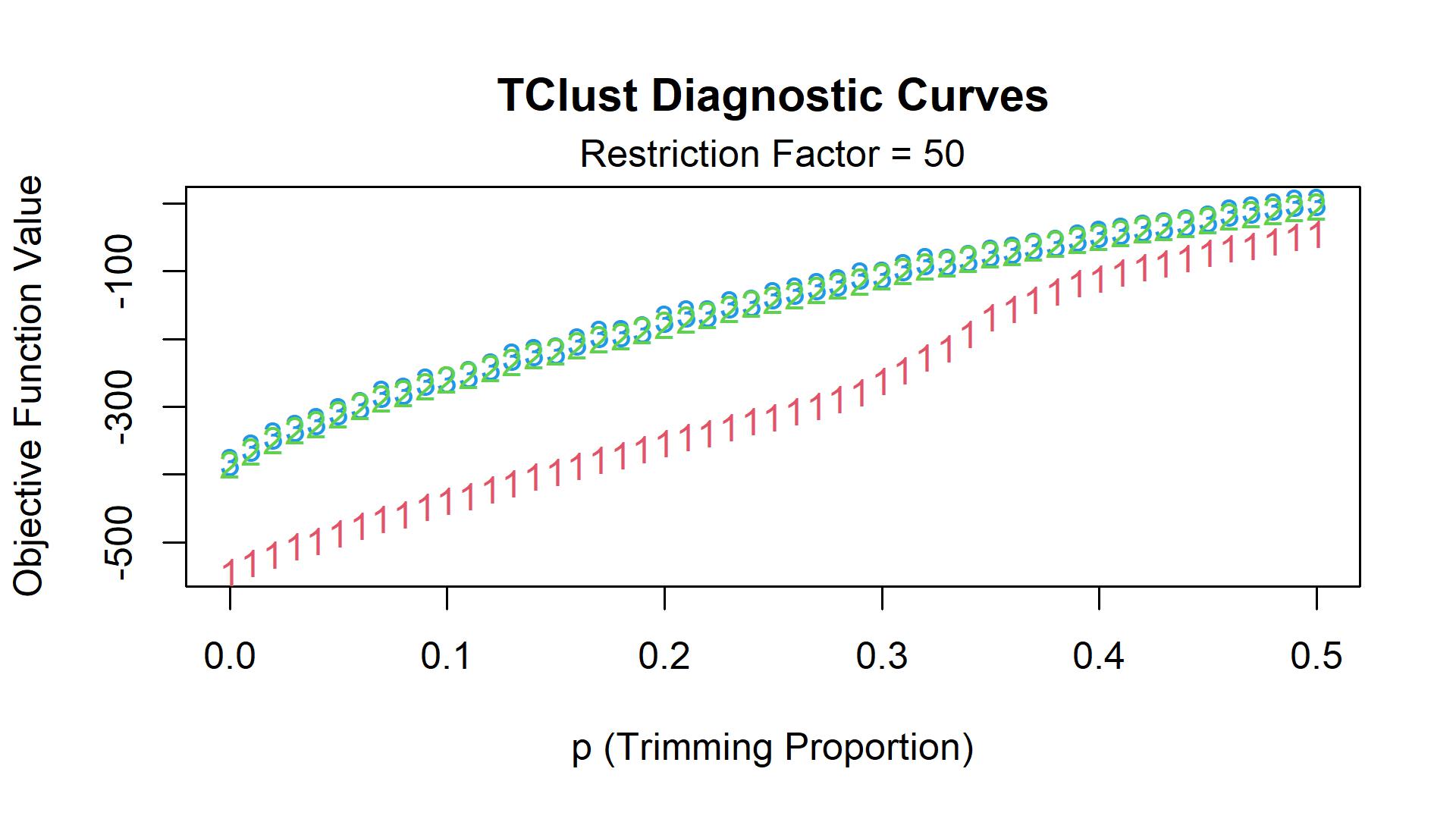}
    \caption{Visualization of the selection tool for Trimmed K-means (\texttt{ctlcurves} routine in R's package \texttt{TClust}). The restriction factor bounds the ratio of maximum to minimum eigenvalues of the covariance matrix of the data, preventing degenerate solutions (set to 50 as standard).}
    \label{figs2:tkm}
\end{figure}

\begin{figure}[H]
    \centering
    \includegraphics[width=0.7\linewidth]{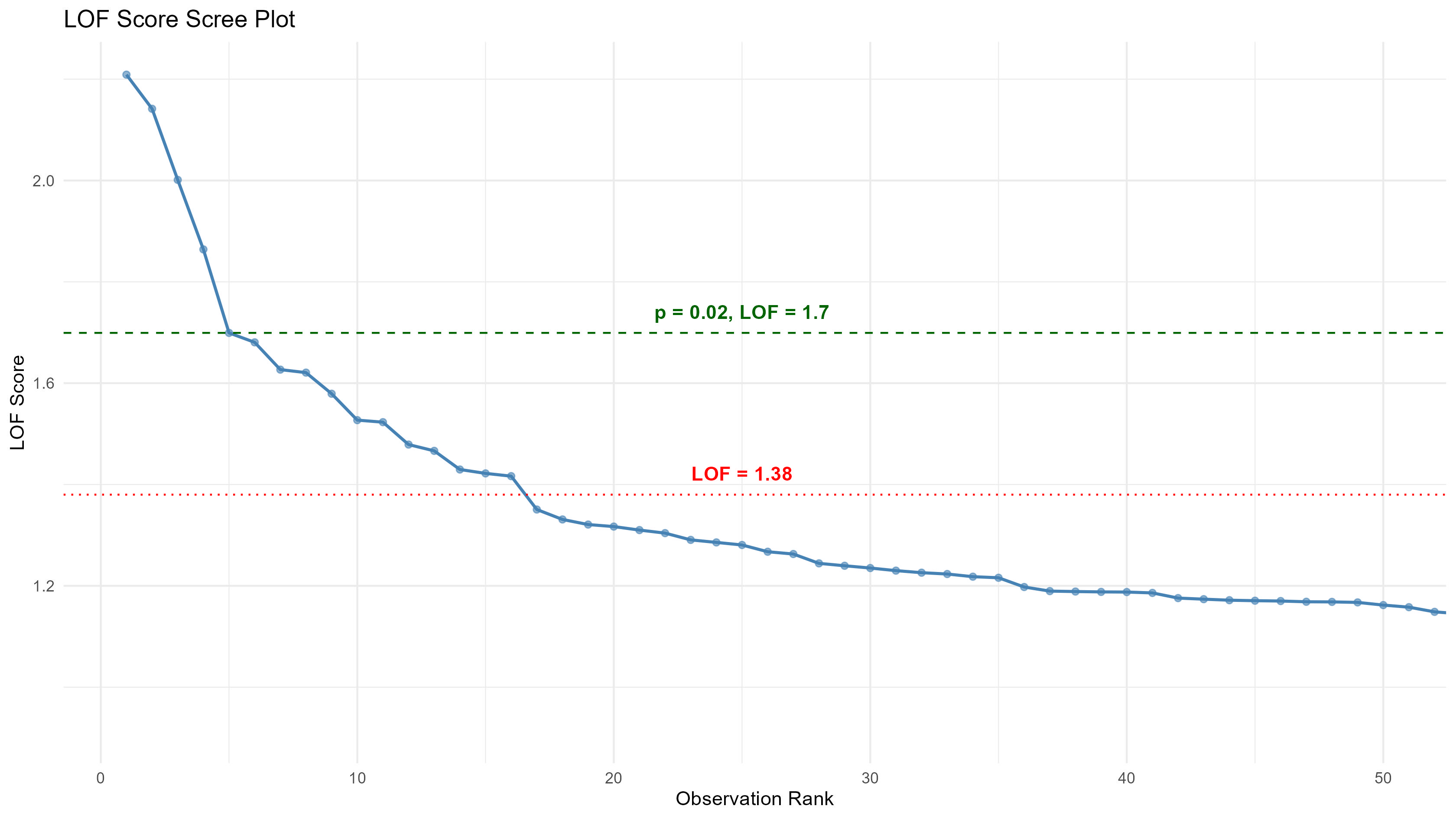}
    \caption{Scree Plot of Observation Rank (increasing order) v. LOF Score.}
    \label{figs3:lof}
\end{figure}

To keep the methodological tutorial in Section 7 as streamlined and readable as possible, we have placed the diagnostic selection plots for Trimmed $K$-means (Figure \ref{figs2:tkm}) and the Local Outlier Factor (Figure \ref{figs3:lof}) in this section.

\noindent For Trimmed $K$-means, the objective function improves sharply moving from one to two clusters, with virtually no improvement at three, confirming the obvious two-cluster structure of the Old Faithful data. However, the objective function improves almost linearly as the trimming proportion increases. Since the dataset lacks major external outliers, the objective function is too robust to show a clear "elbow" for trimming. To enable a direct comparison, we set $\hat{p}=0.018$ (and similarly, $\text{LOF}=1.7$) specifically to match the exact 5 anomalies flagged by ODK-means. In real-world applications where ground truth is unknown, this lack of a clear signal makes tuning $\hat{p}$ much more difficult. 

The LOF diagnostic curve reveals an even bigger practical challenge. The curve displays a prominent elbow at $\text{LOF}=1.38$. Accepting this elbow would classify 16 observations as anomalies, an implausibly high count given the modest sample size ($(N,J)=(272,2)$) and clean distribution of the Faithful data. This contrast highlights a major limitation of traditional methods: fixed trimming quotas and manual cutoffs are hard to tune and often lead to counterintuitive results, hence the need of an adaptive thresholding approach like ODK-means.

\backmatter
\bibliography{sn-bibliography}

\end{document}